\documentclass[aps,prb,groupedaddress]{revtex4-2}

\usepackage {amssymb}
\usepackage {amsmath}
\usepackage{graphicx}

\begin{document}

%Title of paper
\title{Green functions  for a chain subjected to a uniformly varying field in the context of electron transmission}
\author{Lyuba Malysheva}
\email[]{malysh@bitp.kiev.ua}
\affiliation{Bogolyubov Institute for Theoretical Physics, 03680 Kiev, Ukraine}

\date{\today}

\begin{abstract}
On the basis of the tight-binding formalism and Green function 
technique
we obtain all the Green functions matrix elements  for a biased chain with a linear variation of the electron on-site energy.
Their dependence on the system parameters is analyzed in the context of through-molecule electron transport.
\end{abstract}
 \maketitle 
 
 \section{Introduction\label{Sec1}}
 
 During  last decades, fabrication of different molecular contacts of the type "metal-molecular system-metal"   has received an impressive experimental development  allowing high precision measurements of electrical current through single molecules, nanotubes, self-assembled monolayers, nanometer-size dielectric and semiconductor films. However, interpretation of these experiments using  Landauer concept of conductance requires, as a rule, the use of computational simulations of molecular electronic structure. Such studies often give quite limited and method-dependent information, which stimulates the development of   analytical approaches to investigation of electrical properties of molecular contacts.
  
 In the present report, we use the derivation of the transmission coefficient $T(E,V)$, the ratio of the transmitted to incident electron flux at the given energy $E$ and applied voltage $V$, in terms of the coupling function matrix \cite{Mujica,Dat1995,Conn,Chap,Nit2001,Dat2005}. This matrix is determined by the ideal lead Green's functions and molecule-lead interaction matrix. We use the  formulation used in \cite{Chap} whose benefits make it possible to describe molecular contacts  exactly and with the use of realistic model Hamiltonians. 
 Thus, we use the exact analytical expression of the transmission coefficient for a
three-dimensional 3D lead modeled by a cubic semiinfinite
lattice with an arbitrary number of atoms in the surface
and subsurface layers interacting with the molecule \cite{Conn}.
 
In 1960, Wannier introduced the concept of electron energy quantization in solids subjected to a constant homogeneous electric field \cite{Wan1,Wan2}. Actually, his concept was formulated for an infinite monoatomic chain described in the Wannier tight-binding approximation. It can be considered as the theory of Stark effect for a chain of interacting single-level atoms. Therefore, the obtained electron spectrum was named a Wannier-Stark ladder or WS quantization of electron energy, $E_\mu=\mu\varepsilon$, $\mu$ is an integer. The field parameter $\varepsilon$ ($-\varepsilon$) determines the change of the electron potential energy from one atom to the next along (against) the field.

A number of accurate explicit expressions showing 
the electric-field effects on the chain  
electron spectrum have been derived in \cite{Chap,S-G,Fuku,YaG,GV,Davison,SSC,prb63,prb64,WSL}.
The polynomial representation of the exact solution of the spectral problem for the field-affected $\cal N$-atom long tight-binding chain  \cite{Lyuba} was obtained  in the context of through-molecule transport. 

 In this paper, we report  all explicit expressions for the matrix elements of the Green functions for a biased chain with a linear variation of the electron on-site energy
 derived from the exact characteristic equation of Hamiltonian matrix for  ${\cal N}$-length atomic chain. 
 The obtained results are  used for obtaining an
explicit expression of the transmission coefficient of electrons
  through a
spatially finite tilted band. It reveals the resonance structure
of the transmission spectrum and its dependence on the characteristic
parameters of the system. 
   
   \section{Transmission Coefficient}\label{Sec2}

\begin{figure}[t!]
	\centering
		\includegraphics[width=0.68\textwidth]{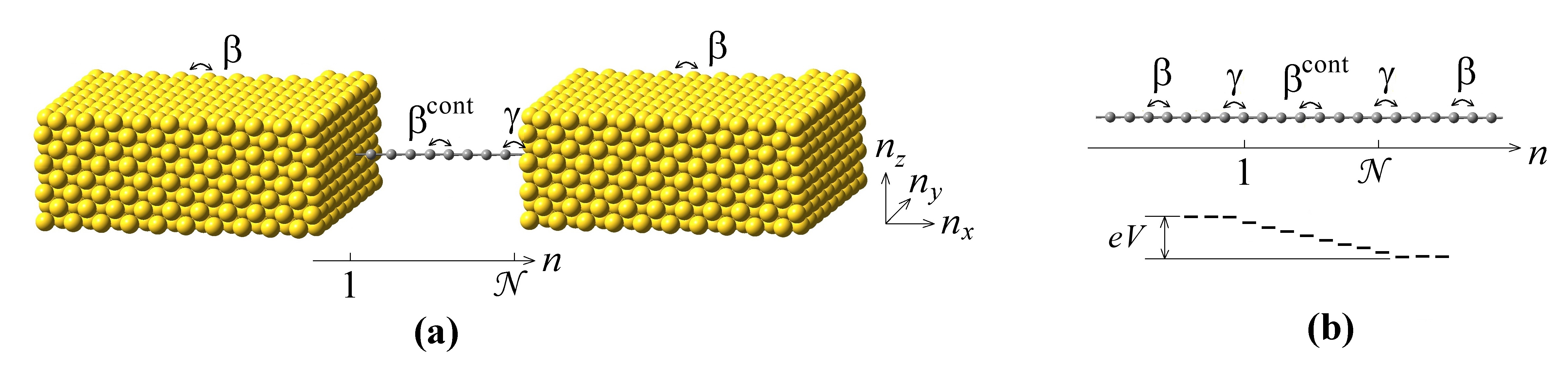}
 \caption{(a). Fragments  of semi-infinite (in $n_x$ direction) left and right leads with a molecular chain in between.
 The binding atoms of the chain are in on-top position. Energies of electron transfer between adjacent atoms
 are $\beta$ in the leads, $\beta^{\rm cont}$ in the molecular chain, and $\gamma$ on the  electrode-molecule interface. (b) One-dimensional case, $N_y=N_z=1$.}
\label{Fig1}
 \end{figure}

Consider a metal wire interrupted by a scattering region but otherwise ideal in the sense that in the absence of the imperfection, electrons could flow freely along the wire. Assume that, as sketched in Fig.~\ref{Fig1}a, it is a molecule, which is coupled in some way to the left and right parts of the wire (the left and right leads), that plays the role of the imperfection. 
In the framework of the Landauer-B{\"u}ttiker theory \cite{Landauer,But,Imry}
 the transmission probability is directly related to the
current-voltage relation.  For the efficient computation and analytic
analysis of the transmission coefficient, the Green
function technique is known to be particularly useful for the
development of efficient computational schemes.
In work \cite{Caroli} it was proposed to describe the
tunnel current in metal-insulator-metal heterostructures
using the Green function language. Later on their treatment has
been reformulated in a number of physical contexts to examine,
in particular, the quantum conductance of molecular
wires \cite{S-G,Fuku,YaG,GV,Davison,Conn}.

In the framework of the Green function formalism,
$T(E,V)$ can be conveniently expressed in terms of the Green
functions referring to the noninteracting left and right leads
and the scattering region. 
 To find $T(E  ,eV)$, we concretize our model as follows. In the bra-ket notation, 
$|{\bf n}\rangle \equiv a^+_{\bf n}|0\rangle = a^+_{n_x}a^+_\perp|0\rangle$ =
$a^+_{n_x}a^+_{n_y}a^+_{n_z}|0\rangle$,
$\langle{\bf n}|{\bf n}'\rangle = \delta_{{\bf n},{\bf n}'}$, the  Hamiltonian of the system "left lead - molecular contact - right lead" depicted in Fig.~\ref{Fig1}
reads
\begin{equation}\label{1}
\hat H = 
\hat H^{\rm L}+\hat H^{\rm cont}+\hat H^{\rm R}+\hat H^{\rm int}.
\end{equation}
Figure~\ref{Fig1} explains the model parameters and shows the  potential profile on the electron way from left  to right electrodes. 
We assume that in the absence of the interaction
between the left/right leads and the contact the eigenstates $\Psi^\mu$ of
the Hamiltonian operator $\hat H^\mu$ of the leads and the contact
($\mu={\rm L}, {\rm R}$, and cont, respectively) can be expanded in a
series of the respective basis set of atomic orbitals $\Psi^\mu=\sum_{{\mathbf n}\in{\mathbf n}_\mu}
\psi^\mu_{\mathbf n} |{\mathbf n}\rangle$. We also treat Hamiltonians $\hat H^{\rm L}$, $\hat H^{\rm R}$ which describe the leads,
as free electron Hamiltonians of semi-infinite cubic
lattices with the electron on-site energy $\varepsilon$ and   the  hopping integral between
 the nearest-neighbor  atoms denoted by $-\beta$ ($\beta>0$). 
 Thus, the energy of transmitted waves is
\begin{equation}\label{2}
E(k_{j_y,j_z})=6\beta-2\beta\left[\cos(k_{j_y,j_z})+\cos\left( \xi_y\right) +
\cos ( \xi_z )\right],\quad
\xi_{(y,z)}\equiv \frac{\pi j_{(y,z)}
}{N_{(y,z)}+1}\quad j_{(y,z)}=1,\dots,N_{(y,z)}, 
\end{equation}
where   $k_{j_y,j_z}$ is real (imaginary) for propagating (evanescent)
 modes.

The left-to-right drop of the applied potential $eV$ is taken into
account as a  shift of the site energies of each atom by the field parameter $\varepsilon_F=eFa/\beta$, where
$e$ is the absolute value of electron charge, $F$ is the electric field strength, and $a$ is the lattice
constant.  Thus, it is assumed that the potential difference between the left  and right electrodes drops linearly inside the contact:
$eV=\varepsilon_F({\cal N}+1)$, 
and the Hamiltonian of molecular chain has the form
%3
\begin{equation}\label{3}
\hat H^{\rm cont} = \sum_{n  =1}^{\cal N} 
\left\{(\varepsilon^{\rm cont}-n  \varepsilon_F)a^+_{n  }a_{n  } 
- \beta^{\rm cont}\big[(1-\delta_{n  ,1})a^+_{n  -1}+
(1-\delta_{n  ,{\cal N}})a^+_{n  +1}\big ]a_{n  }\right\}.
\end{equation}
We consider a simplified model of metal-molecular
interaction which involves only two of all molecule
atoms: these binding atoms have coordinates $n=1$ and $n=\cal N$.
The interaction operator is then given by 
%4
\begin{equation}\label{4}
\hat H^{\rm int}=\gamma \sum_{{\mathbf n}\in{\mathbf n}_{\rm L}}|1\rangle \langle {\mathbf n}|
+\gamma \sum_{{\mathbf n}\in{\mathbf n}_{\rm R}}|{\cal N}\rangle \langle {\mathbf n}|,
\end{equation}
i.e., the parameter 
$\gamma$ accounts to the difference between the electron transfer rates from contact to electrodes and backward. 

By definition, the transmission coefficient is equal to the ratio of the transmitted electron flux 
  to the incident flux. 
The derivation of the
transmission coefficient is well known \cite{Mujica,Dat1995,Conn,Chap}. Due to simplifying
model assumptions, this principal quantity can be obtained  in
a fully analytical form via solving the Lippman-Schwinger equation with the Hamiltonian $\hat H$.
Here, we use $T(E  ,eV)$ in the following form:
%5
\begin{equation}\label{5}
T(E  ,eV)=4{\rm Im}(A^{\rm L}){\rm Im}(A^{\rm R})
\left |  \dfrac{G^{\mathstrut \rm cont}_{1,{\cal N} }}{\left (1-A^{\rm L}G^{\mathstrut \rm cont}_{1,1 }\right )\left (1-A^{\rm R}G^{\mathstrut \rm cont}_{{\cal N},{\cal N} }\right ) -A^{\rm L}A^{\rm R}(G^{\mathstrut \rm cont}_{1,{\cal N} })^2 } \right |^2,
\end{equation}
where $G^{\mathstrut \rm cont}_{n  ,n'  }$ are the Green functions for the Hamiltonian (\ref{3})  and the coupling functions $A^{\rm L}, A^{\rm R}$ are defined as follows:
%6
\begin{equation}\label{6}
\begin{split}
A^{({\rm L},{\rm R})}&=\gamma^2 \sum_{{\mathbf n},{\mathbf n'}\in{\mathbf n}_{({\rm L},{\rm R})}}G^{({\rm L},{\rm R})}_{{\mathbf n},{\mathbf n'}}, \\ 
G^{({\rm L},{\rm R})}_{1,n_y,n_z;1,n_y',n_z'}&=-\dfrac{1}{\beta}\dfrac{4}{(N_y+1)(N_z+1)}\sum_{j_y=1}^{N_y}\sum_{j_z=1}^{N_z}
e^{ik^{({\rm L},{\rm R})}_{j_y,j_z} }\sin(\xi_y n_y)\sin(\xi_z n_z)\sin(\xi_y n_y')\sin(\xi_z n_z').
\end{split}
\end{equation}
Finding the transmission coefficient for the model specified by the Hamiltonian operator in Eq.~(\ref{3}) requires the knowledge of the Green's function matrix elements appeared in Eq.~(\ref{5}). They are found in the next section.
From now and on the model parameters $\varepsilon$, 
$\varepsilon^{\rm cont}$, $\beta^{\rm cont}$, $\gamma$, and
$\varepsilon_F$ will be expressed using $\beta$   as the unit of energy.

\section{Green's functions for a tilted chain in the framework of the tight-binding model }\label{Sec3}

The system of equations for finding the required   Green's functions reads
%7
\begin{equation}\label{7}
\left(E  -\varepsilon^{\rm cont}+n   \varepsilon_F\right)G^{\mathstrut \rm cont}_{n  ,n'  }
=-\beta^{\rm cont}[G^{\mathstrut\rm cont}_{n  -1,n  '}(1-\delta_{n  ,1})
+G^{\mathstrut\rm cont}_{n  +1,n  '}(1-\delta_{n  ,N})]+\delta_{n  ,n  '}, \quad n  ,n  '=1,\dots,{\cal N}.
\end{equation}
To find all the matrix elements $G^{\mathstrut \rm cont}_{n  ,n'  }$, it is convenient to use the generating functions method.
We define the generating function as follows:
%8
\begin{equation}\label{8}
{\cal G}_{n'}(\varphi)=
\sum_{n=1}^{\cal N} G^{\mathstrut\rm cont}_{n  ,n'  }(E,eV) e^{in \varphi}.
\end{equation}
It can be verified by direct substitution that ${\cal G}(\varphi)$ satisfies the following differential equation:
%9
\begin{equation}\label{9}
(E  - \varepsilon^{\rm cont}-2\beta^{\rm cont}\cos\varphi)
{\cal G}_{n'}(\varphi) + i \dfrac{\beta^{\rm cont}}{a{\cal F}}
\frac{d{\cal G}_{n'}(\varphi)}{d\varphi}=
e^{in  ' \varphi}-G^{\mathstrut\rm cont}_{1,n  ' }(E,eV)
-e^{i ({\cal N} +1)\varphi} G^{\mathstrut\rm cont}_{{\cal N},n  '}(E,eV). 
\end{equation}
Solving this linear differential equation allows to express the solution of (\ref{7}) in terms of Bessel functions of the first and second kind:
%10
\begin{equation}\label{10}
\begin{split}
&G^{\mathstrut\rm cont}_{n  ,n'  }(E,eV)= -\frac{\pi}{ \varepsilon_F
\big[J_{\nu}(z)Y_{\nu-{\cal N}-1}(z)-J_{\nu-{\cal N}-1}(z)Y_{\nu}(z)\big]}
\\[20pt]
&\times
\left \{
\begin{array}{l} 
  \big[J_{\nu-n  }(z)Y_{\nu-{\cal N}-1}(z)-J_{\nu-{\cal N}-1}(z)Y_{\nu-n  }(z)\big]  
\big[J_{\nu}(z)Y_{\nu-n'  }(z)-J_{\nu-n  '}(z)Y_{\nu}(z)\big],
\quad n   \ge n'  ,
\\ \\
n   \leftrightarrow n'  ,\quad   n   \le n'  , \quad
z = \dfrac{2\beta^{\rm cont}}{ \varepsilon_F},\quad 
\nu=-(E  - \varepsilon^{\rm cont}) \dfrac{{\cal N}+1}{eV} = -(E  - \varepsilon^{\rm cont})
\dfrac{1}{ \varepsilon_F}.
\\
\end{array}
\right.
 \end{split}
\end{equation}
For the particular values of $n  , n'  =1,{\cal N}$, Eq. (\ref{10}) gives the expressions for the Green's functions used in the definition of the transmission coefficient $T(E,eV)$:
%11
\begin{align}\label{11}
{\cal D}_G(E,eV)G^{\mathstrut\rm cont}_{1,1}(E,eV)& = 
J_{\nu - 1}(z)Y_{\nu -{\cal N}- 1}(z) - 
Y_{\nu - 1}(z) J_{\nu -{\cal N}- 1}(z)\equiv \tilde{G}_{1,1} , \notag
\\
{\cal D}_G(E,eV)G^{\mathstrut\rm cont}_{{\cal N},{\cal N}}(E,eV)&
=J_{\nu}(z)Y_{\nu - {\cal N}}(z) - Y_{\nu}(z)J_{\nu - {\cal N}}(z)\equiv \tilde{G}_{{\cal N},{\cal N}},\notag \\
{\cal D}_G(E,eV){G}^{\mathstrut\rm cont}_{1,{\cal N}}(E,eV)&=\dfrac{1}{\pi}
\dfrac{\varepsilon_F}{\beta^{\rm cont}}\equiv \tilde{G}_{1,{\cal N}},\notag\\
Q(E,eV)&={\cal D}_G(E,eV)\left \{  {G}^{\mathstrut\rm cont}_{1,1}(E,eV){G}^{\mathstrut\rm cont}_{{\cal N},{\cal N}}(E,eV)-\big[{G}^{\mathstrut\rm cont}_{1,{\cal N}}(E,eV) \big ]^2\right \}\notag
\\&=\dfrac{1}{\beta^{\rm cont}} \left[
 J_{\nu - 1}(z) Y_{\nu - {\cal N}}(z) - Y_{\nu - 1}(z) J_{\nu - {\cal N}}(z)
 \right],
\end{align} 
and 
%12
\begin{equation}\label{12}
{\cal D}_G(E,eV) =-\beta^{\rm cont} \left[
J_{\nu}(z)Y_{\nu - {\cal N} - 1}(z)-Y_{\nu}(z)J_{\nu - {\cal N} - 1}(z)\right] .
\end{equation} 

Note that for the case ${\cal N}=1$, making use of the well-known relations for the Bessel functions 
%13
\begin{equation}\label{13}
\begin{split}
J_{\nu-1}(z)Y_{\nu-2}(z)&-Y_{\nu-1}(z)  J_{\nu-2}(z)=
\dfrac{2}{\pi z},\\
J_{\nu-1}(z)+J_{\nu+1}(z)&=\dfrac{2\nu}{z}J_{\nu}(z), \quad
Y_{\nu-1}(z)+Y_{\nu+1}(z)=\dfrac{2\nu}{z}Y_{\nu}(z),
\end{split}
\end{equation}
in Eqs.~(\ref{11}), (\ref{12}), we obtain the evident relation
%14
\begin{equation}\label{14}
G^{\mathstrut\rm cont}_{1,1}(E,eV) = \dfrac{1}{E   - ( \varepsilon^{\rm cont} - eV/2)} .
\end{equation}

Relations (\ref{10}) give analytical expressions for the Green 
functions for a biased linear chan which can be used for analytical modeling in a great number of 
applications. Substituting Eqs. (\ref{11}) in Eq. (\ref{5}) allows to find the transmission coefficient
for the system depicted in Fig.~\ref{Fig1}. In the next section, we use these results to derive $T(E,eV) $
for the case $N_y=N_z=1$, i.e., for an atomic chain shown in Fig.~\ref{Fig1}b.

\section{One-dimensional case}

For the case $N_y=N_z=1$, our model   corresponds to an atomic chain shown in Fig.~\ref{Fig1}b. The site energy along the chain equals 2 (to recall, in $\beta$ units) for $n_x\leq 0$, $2-eV$ for $n_x> {\cal N}$, and $2 -\varepsilon_F n$, $\varepsilon_F= eV/(N+1)$ for $n\in\overline{1,\cal N}$ in, respectively, the left and right electrodes and contact. 
The eigen energies (\ref{2}) for Hamiltonian (\ref{1}) simplify in this case to
%15
\begin{equation}\label{15}
E=2\left(1 - \cos k^{\rm L}\right)=2\left(1 - \cos k^{\rm R}\right) - eV,
\quad 0\le k^{\rm L},k^{\rm R}\le\pi,
\end{equation}
with the wave vectors in units of the inverse interatomic distance $a^{-1}$. Relation (\ref{5}) is rewritten as follows:
\begin{equation}\label{16}
T (E ,eV )=
\dfrac{4\gamma^4\sin k^{\rm L}\sin k^{\rm R}(G^{\mathstrut\rm cont}_{1,{\cal N}})^2}
{\left|1+\gamma^2\left(e^{ik^{\rm L}}G^{\mathstrut\rm cont}_{1,1}+e^{ik^{\rm R}}G^{\mathstrut\rm cont}_{{\cal N},{\cal N}}\right) 
+\gamma^4e^{i(k^{\rm L} + k^{\rm R})}
\left[G^{\mathstrut\rm cont}_{1,1} G^{\mathstrut\rm cont}_{{\cal N},{\cal N}}-\left (G^{\mathstrut\rm cont}_{1,{\cal N}}\right )^2\right]\right|^2}.
\end{equation}
Using relations for the Green functions (\ref{11}), after some algebra, we get
%17
\begin{align}\label{17}
{\cal D}_{T}T(E,eV)&=
4\gamma^4\sin k^{\rm L}\sin k^{\rm R}
 \tilde{G}_{1,{\cal N}}^2, \notag \\
{\cal D}_T &\equiv  \Big [ {\cal D}_G + 
\gamma^2\left(\cos k^{\rm L}\tilde{G}_{1,1} + 
\cos k^{\rm R}\tilde{G}_{{\cal N},{\cal N}}\right) + \gamma^4
\cos\left(k^{\rm L}-k^{\rm R}\right)Q(E,eV)\Big ]^2 \notag \\
&\quad+ \gamma^4\Big [\sin k^{\rm L} \tilde{G}_{1,1}
 - \sin k^{\rm R}\tilde{G}_{{\cal N},{\cal N}} +\gamma^2
\sin\left(k^{\rm L} - k^{\rm R}\right)Q(E,eV)\Big ]^2
+4\gamma^4\sin k^{\rm L}\sin k^{\rm R}\tilde{G}^2_{1,{\cal N}}.
\end{align}
The condition of transmission without backscattering, $T(E,eV) = 1$, directly follows from Eq.~(\ref{17}):
%18 
\begin{align}\label{18}
 \Big [\gamma^{-2} {\cal D}_G + 
\cos k^{\rm L}\tilde{G}_{1,1} &+ \cos k^{\rm R}\tilde{G}_{{\cal N},{\cal N}}
 + \gamma^2
\cos\left(k^{\rm L} - k^{\rm R}\right)Q(E,eV)\Big ]^2 \notag \\
&+ \Big [\sin k^{\rm L} \tilde{G}_{1,1} - \sin k^{\rm R}
\tilde{G}_{{\cal N},{\cal N}} +\gamma^2
\sin\left(k^{\rm L} - k^{\rm R}\right)Q(E,eV)\Big]^2 
=0.
\end{align}
In the particular case,  $\beta^{\rm cont} = 1$ and $eV = 0$ (and, consequently, $k^{\rm L}= k^{\rm R}=k$), Eq.~(\ref{17}),  repeats the result obtained earlier \cite{Chap}: 
%19
\begin{equation}\label{19}
T(E  ,0)=\dfrac{4\gamma^4\sin^4k  }
{\big\{[\sin({\cal N}+1)k  ]-2\gamma^2\cos k  \sin {\cal N}k  +\gamma^4\sin[({\cal N}-1)k  ]\big\}^2
+4\gamma^4\sin^4k  }. 
\end{equation}

Expression (\ref{17}), though much simpler than Eq.(\ref{5}), still remains fairly complicated.
However,  for the case of small potential difference, $eV<<1$, it can be significantly simplified
 with the help of the approximation similar to that used in \cite{prb64}. 
Namely, for any coupling parameter $\gamma$ and any $\cal N$, the transmission coefficient $T(E,eV)$ can be approximated as follows:
%20
\begin{equation}\label{20}
T(E,eV)\approx\dfrac{4\gamma^4}
{4\gamma^4+(1 - \gamma^4)^2\sin^2\left[p+\dfrac{{\cal N} + 1}{2}
\left(E - 2 + \dfrac{eV}{2}\right)\right]}, \quad p\equiv \dfrac{\pi}{4}[(-1)^{\cal N} + 1].
\end{equation}

\begin{figure}[t!]
	\centering
		\includegraphics[width=0.95\textwidth]{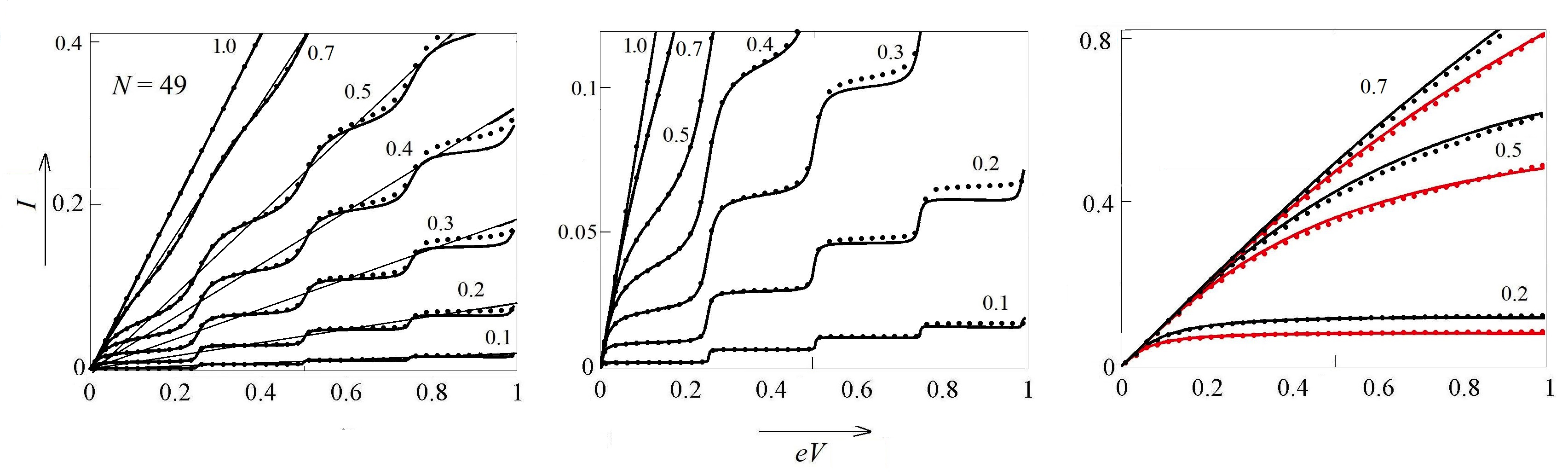}
 \caption{Left panel: The exact dependence $I(eV)$, Eq.~(\ref{21}) -- solid lines, and its  approximation~(\ref{22}) -- dotted lines, calculated for $N=49$ and $\gamma=0.1, 0.2, 0.3, 0.4, 0.5, 0.7, 1.0$. Central panel: Enlarged  dependences shown on the left panel. Right panel: The same dependences for $N=3$ (black), $N=5$ (red) and $\gamma=0.2,  0.5$, and 0.7.}
\label{Fig2}
 \end{figure}

Following the Landauer-B{\"u}ttiker theory \cite{Landauer,But},
we express the
current-voltage relation  via  the transmission coefficient in the form
\begin{equation}\label{21}
I(eV) = \dfrac{2e}{h}
\int\limits_{\max\left(0,2-eV \right)}^{\min\left(2,4-eV\right)}T(E,eV)dE.
\end{equation}
The limits of integration $[2 - eV,2]$, $0 \le eV \le 2$ and $[0,4 - eV]$, $2 \le eV \le 4$ correspond to the nonzero values of the transmission coefficient dictated by the Pauli exclusion principle. 
Since, for the condition $eV<<1$,  the obtained approximation for the transmission coefficient (\ref{20}) admits the exact integration, we derive an explicit expression for the current as a function of the potential difference, $\gamma$, and  $\cal N$:
%22
\begin{equation}\label{22}
I(eV) = \dfrac{2e}{h} \dfrac{8\gamma^2}{({\cal N}+1)(1+\gamma^4)}\arctan
\left \{\dfrac{1+\gamma^4}{2\gamma^2}\tan\left [p+ \dfrac{({\cal N}+1)eV}{4}\right ]\right\},\quad 
eV\lesssim1.
\end{equation}
As illustrated in Fig.~\ref{Fig2}, this approximation works reasonably well in the interval $0<eV\lesssim 1$.
 Thus, simple analytic approximation (\ref{22}) satisfactory reproduces the current-voltage dependences for molecular contact depicted in Fig.~\ref{Fig1} as functions of the coupling parameter and the contact length for the potential differences varying from zero to several electron-volts.

\end{document}